\documentclass[5p,4pt,preprint]{elsarticle}

\usepackage{amssymb}

\usepackage{txfonts}
\usepackage{graphicx}
\usepackage{amssymb}
\usepackage{natbib}

\newcommand{\eqb}{\begin{eqnarray}}
\newcommand{\eqe}{\end{eqnarray}}
\newcommand{\sth}{\sigma_{\rm T}}
\newcommand{\gpmn}{\gamma_{\rm p,min}}
\newcommand{\gpmx}{\gamma_{\rm p,max}}
\newcommand{\gemn}{\gamma_{\rm e,min}}
\newcommand{\gemx}{\gamma_{\rm e,max}}
\newcommand{\tpesc}{t_{\rm p,esc}}
\newcommand{\teesc}{t_{\rm e,esc}}

\def\apj{ApJ}%
\def\apjl{ApJ}%
\def\aap{A\&A}%
\def\mnras{MNRAS}%
\def\prd{Phys.~Rev.~D}%
\def\nat{Nature}%
\journal{Astroparticle Physics}

\begin{document}

\begin{frontmatter}

\title{Self-consistent neutrino and UHE cosmic ray spectra from Mrk~421}

\author[uoa]{Stavros Dimitrakoudis\corref{cor2}}
\ead{sdimis@phys.uoa.gr}

\author[uoa]{Maria Petropoulou\corref{cor1}}
\ead{maroulaaki@gmail.com}

\author[uoa]{Apostolos Mastichiadis\corref{cor1}}
\ead{amastich@phys.uoa.gr}

\address[uoa]{Department of Physics, University of Athens, Panepistimiopolis, GR 15783 Zografos, Greece}

\cortext[cor1]{Corresponding author}
\cortext[cor2]{Principal corresponding author}

\begin{abstract}
We examine the neutrino and cosmic ray spectra resulting from two models
of fitting the spectral energy distribution (SED) of the blazar
Mrk~421 using a self-consistent leptohadronic code.
The $\gamma$-ray emission is attributed to either
synchrotron radiation of ultra-high energy protons (LHs model)
or to synchrotron radiation from
electrons that result from photopion interactions of lower energy
protons (LH$\pi$ model).
Although both models succeed in fitting
satisfactorily the SED, the parameter values that they use 
result in significantly 
different neutrino and cosmic-ray spectra.
For the LH$\pi$ model, which requires high proton energy density, we find that
the neutrino spectrum peaks at an energy $E_{\rm \nu,peak}= 3.3$ PeV which falls well
within the energy range of recent neutrino observations. While at the same time its peak
flux is just under the sensitivity limit of IC-40 observations, it cannot
produce ultra-high energy cosmic rays. In the LHs model, on the other hand,
neutrinos are far from being detectable because
of their low flux and peak energy at $E_{{\rm \nu,peak}} \simeq 100$ PeV.
However, the propagation of protons produced by the decay of escaping
neutrons results in an ultra-high energy cosmic ray flux close to that
observed by Pierre Augere, HiRes and Telescope Array at energies 
$E_p \simeq 30$ EeV.

\end{abstract}


\end{frontmatter}

\section{Introduction}
\label{intro}

Fermi and ground-based TeV observations have convincingly proved that
blazars, a class of Active Galactic Nuclei having their jets pointed towards us, 
can be considered as very efficient particle accelerators.
It is possible then that an active region in a blazar jet, such as a standing shockwave, can 
accelerate both protons and electrons to high energies.
In such a case electron synchrotron radiation is responsible
for the blazar emission from the radio up to UV/X-ray regime 
while the high energy (GeV-TeV) emission
is attributed to processes induced by hadrons which include proton
synchrotron radiation \citep{Mucke2001,aharonian00} and pion-related
cascades \cite{mannheimbiermann92}, producing thus the 
characteristic double hump
blazar Spectral Energy Distribution (SED) in a $\nu F_\nu$ vs. $\nu$ diagram.

It is well known that the ultimate proof for the
existence of high energy protons in
blazar jets can come only 
from the detection of high energy neutrinos.
An accurate modelling of such neutrino spectra is vital for the 
interpretation of observations by neutrino telescopes, 
 especially 
in the context of the recent observation of neutrino showers in the PeV energy range by IceCube \citep{IcePeV2013}.
A common approach is to suppose a generic proton distribution at
the source and from that to obtain a neutrino spectrum, which can
then be integrated over redshift to give us the total flux at Earth
 (e.g., \citep{Rachen1998, Mannheim2001}, or in the context of PeV neutrinos,
  \citep{Kistler2013}).

In the present paper we will follow a different approach. 
We will use a recently developed numerical code 
that treats the radiative transfer problem in a region
where both electrons and protons are accelerated. 
This  will allow us to calculate
the proton distribution at source simultaneously with the
radiated photon spectrum. 
We will consequently
seek successful fits to the SED of the well monitored
BL Lac object Mrk~421. 
As the numerical code calculates not only the photon spectra
but also the emergent neutron and neutrino ones,
the obtained SED fits will automatically give the production
spectra of these two other species expected from the particular
source.

In Section \ref{model} we briefly present the numerical
code and the relevant free parameters. In Section \ref{results} we 
present the results for two applications of the model, and we conclude
in Section \ref{discussion} with a summary and some discussion.

\section{The model}
\label{model} 

In this section we will present the principles of the adopted
numerical code. 
While details can be found in \citep{DMPR2012},  for the
sake of completeness we summarise here its basic points.
We have assumed that protons and electrons, 
accelerated\footnote{In the present work we have not
specified the acceleration mechanism that pushes both protons and electrons to ultra-high energies; the maximum Lorentz factor and 
the slope of the 
accelerated particle distributions are rather treated as free parameters that are solely determined by the
fit of the SED. The only restriction we impose is the Hillas criterion.} up to high energies,
are injected uniformly 
with a constant rate inside a spherical
volume of radius R containing a tangled magnetic field of strength B. Both species
lose energy by various physical processes: 
Protons lose energy by 
synchrotron radiation, photopair and photopion production. 
All photopion interactions,
including the decay of produced pions and muons,
have been modelled using the results of the SOPHIA event generator 
\citep{SOPHIA2000}, while photopair interactions are modelled using 
the Monte Carlo results of \citep{Protheroe1996}. 
The stable secondary 
products of these processes include electron/positron pairs 
(which will be hereafter collectively referred to as electrons), photons,
neutrons and neutrinos. Muons, which are unstable 
secondary products, are allowed to lose energy through synchrotron
radiation before their decay.   
Electrons (primary and secondary)  
lose energy by synchrotron
radiation and inverse Compton scattering,
while gamma-rays can be absorbed on the softer photons,
produced mainly by electron synchrotron radiation, creating
more electron-positron pairs. 
Neutrons, not being confined by the magnetic fields in the source, 
can either escape or interact with the photons before decaying back
to protons. Finally neutrinos constitute the most uncomplicated component 
as they escape from the source essentially with their production spectrum.
 
In order to follow the evolution of this 
non-linear (see \citep{PM2012})
system we use the kinetic equation approach;
thus we write five time-dependent equations, for each stable species, namely
for protons, electrons, photons, neutrons and neutrinos.  The various rates
are written in such a way as to ensure self-consistency, i.e. the amount of
energy lost by one species in a particular process is equal to that emitted (or injected)
by another. That way one can keep the logistics of the system in the sense that
at each instant the amount of energy entering the source through the injection 
of protons and primary electrons should equal the amount of energy escaping from it in the form of 
photons, neutrons and neutrinos (in addition to the energy carried away due to electron and proton escape from the source). 
We have not included an additional equation describing the evolution of pions 
 because their decay rate is far
higher than its synchrotron energy loss rate, so the latter can be safely
neglected. As this condition does not apply for muons we have made an adjustment to the code 
described in \citep{DMPR2012}. For each muon energy, we calculate the energy lost
to synchrotron radiation before it decays. Then the secondaries  
corresponding to the reduced muon energy, whose yields have also been 
computed by the SOPHIA event generator \citep{SOPHIA2000}, are essentially produced
instantaneously. Moreover, these products  along
with muon synchrotron photons are added to their respective kinetic
equations without the need for a separate  equation for muons.
    
The above approach has various advantages: since it uses a particle injection rate, 
one can calculate the efficiency of the model -- note that the most usual
approach of using a particle distribution function in order to calculate the resulting
spectra cannot give such estimates. Furthermore, it can calculate in an exact way
-- under the assumption of a spherical geometry --
the photon, neutron and neutrino transport and thus compare directly their respective fluxes. 
Finally it can treat both stationary and time-dependent cases, thus 
providing a tool for studying time variability in the context of a  leptohadronic model -- see \citep{MPD2013}.

Our free parameters, as measured in the rest frame of the emitting region, are limited to the following:

\begin{enumerate}
\item the injected luminosities of
protons and electrons, which are expressed in terms 
of compactnesses:
\eqb
\ell_{\rm i}^{\rm inj}={{L_{\rm i} \sth}\over{4\pi R m_i c^3}},
\label{lpinj}
\eqe
where i denotes protons or electrons and 
$\sigma_{\rm T}$ is the Thomson cross section;

\item the upper Lorentz factors of the injected protons and electrons, 
$\gpmx$ and $\gemx$ respectively; 

\item the power law indices $p_{\rm p}$ and $p_{\rm e}$ of injected
protons and electrons, respectively;

\item the radius R of the region; 

\item its magnetic field, B; 

\item its Doppler factor, $\delta$; and

\item the escape time
for both particles, which is assumed to be the same ($\tpesc=\teesc$).

\end{enumerate}
Finally. the lower Lorentz factors are assumed to be $\gpmn=\gemn=1$. 

As in \citep{MPD2013}, we have adopted a cosmology 
with $\Omega_{\rm m}=0.3$, $\Omega_{\Lambda}=0.7$ and 
$H_0=70$ km s$^{-1}$ Mpc$^{-1}$, where the redshift of Mrk 421 $z=0.031$ 
corresponds to a luminosity distance $D_{\rm L}=0.135$ Gpc.

\section{Results}
\label{results} 
Being one of the closest blazars to Earth, Mrk~421 has been the
target of multiple multiwavelength (MW) campaigns over the years. 
Observations by \citep{Fossati2008} in 2001 produced excellent sets
of time-dependent data both at the X-ray (RXTE) and TeV (Whipple and
HEGRA) regimes during a six-day period. In
 \citep{MPD2013} we used a variation of our model to fit Mrk~421 in 
its preflare state of March 22nd/23rd 2001 and then applied
time-dependent fluctuations of the injection compactnesses or of the maximum
energies of injected protons to study the interplay between 
variations of X-ray and $\gamma-$ray fluxes. 
We corrected also our model predicted $\gamma-$ray spectra for 
photon-photon absorption on the extragalactic background light (EBL) by using
the model of \cite{kneiske04}. However, the effects of EBL absorption are negligible
due to the proximity of the source to Earth.
Our fitting is based only on X-ray and TeV observations since no 
GeV data were
available during the 2001 campaign. For comparison reasons, we 
included in our model SEDs the Fermi data during 
the period of the IceCube 40-string configuration (IC-40)
 \citep{Abdo2011}. Inclusion of the
additional spectral information
would only slightly alter our parameter values. 
Here we have focused 
on the steady state emission of Mrk~421 in order to examine the 
resulting neutrino and
neutron distributions in relation to the observed photon spectra.
Instead of using the conventional IC-40 upper limits \citep{abbasietal11}, which are derived for soft
neutrino spectra emitted by a generic source,
we adopted those  given by \citep{Tchernin2013} that are calculated specifically for Mrk~421 and
for different power-law neutrino spectra.

We note here that for the SED modelling of Mrk~421 we did not attempt 
a best $\chi^2-$ fit but instead 
we followed the approach of \citep{petromast12}, where they considered families 
of acceptable fits to the data. This allows us to use a narrow range of parameter values.
Within this range, we found, interestingly enough, two sets of 
very different proton injection parameters
which give good fits to the data.
In both cases, optical and  X-rays are fitted by the primary
injected leptonic component, while the origin of the GeV-TeV emission is 
different between the two models.
Fitting of the optical data requires in both cases electron
distributions that are flat, still with indices ($p_{\rm e}=1.2$) that are compatible with certain acceleration
models (e.g. \citep{Niemiec2004})\footnote{In the above fits we have used $\gamma_{\rm e, min}=1$.
If we were to relax this constraint and allow higher values of the lower energy cutoff we could obtain
acceptable fits with steeper power-law distributions, which would be easier to reconcile with more  acceleration models.
}.  In what follows, we will discuss the resulting photon, neutrino and ultra-high energy cosmic ray 
(UHECR) spectra obtained in both cases.

\begin{table}
\centering
\begin{tabular}{l  c c}
 \hline 
Parameter & model LH{$\pi$} & model LHs  \\
\hline \hline

$\gpmx$ & $3.2 \times 10^6$ & $6.3 \times 10^9$ \\
$\gemx$ & $8 \times 10^4$ & $4 \times 10^4$ \\
\hline

$p_{\rm p}$ & 1.3 & 1.5 \\
$p_{\rm e}$ & 1.2 & 1.2 \\
\hline

$\ell_{\rm p}$ & $2 \times 10^{-3}$ & $1.6 \times 10^{-7}$ \\
$\ell_{\rm e}$ & $3.2 \times 10^{-5}$ & $ 10^{-4}$ \\
\hline

R (cm) & $3.2 \times 10^{15}$ & $3.2 \times 10^{15}$ \\
B (G) & 5 & 50 \\
$\delta$ & 26.5 & 21.5 \\
\hline

\end{tabular}
\caption{Initial parameters for the two fits.}
\label{table1}
\end{table}

\begin{figure}
\resizebox{1.0\hsize}{!}{\includegraphics{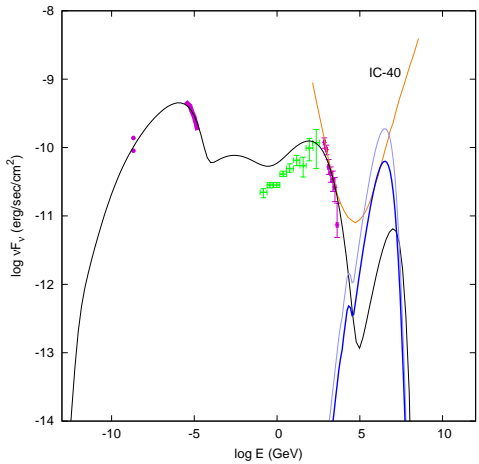}}
\caption{\small{Spectra of photons (black line) fitting the March 22nd/23rd 2001 observation of Mrk~421 (purple points), 
neutrinos of all flavours (light blue line) and muon neutrinos 
(thick blue line)
arriving at Earth, after taking into account neutrino oscillations
 according to the photopion model 
LH{$\pi$}. 
Fermi observations \citep{Abdo2011} (green points) are not simultaneous with the rest of the data. 
The 40-String IceCube limit for muon neutrinos
\citep{Tchernin2013} is plotted with an orange line.
}}
\label{MW_photopion}
\end{figure}
\begin{figure}
\centering
\resizebox{1.0\hsize}{!}{\includegraphics{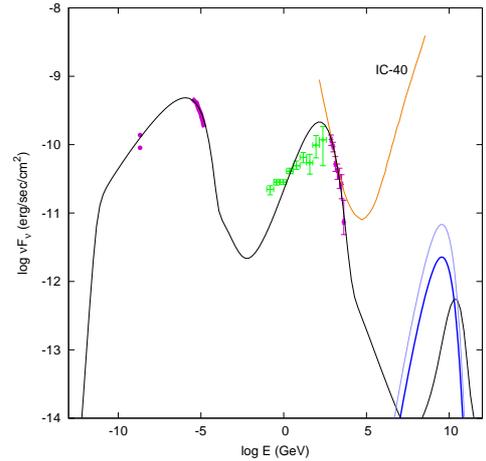}}
 \caption{\small{Same as Fig. \ref{MW_photopion} but for the
proton synchrotron model LHs.
}}
\label{MW_psyn}
\end{figure}
\subsection{Photon emission}
We label as `LH$\pi$' 
(Leptohadronic-pion) and 
`LHs' \\ (Leptohadronic-synchrotron) the two models 
based on the parameter sets mentioned in the previous
section according to the origin of the $\gamma$-ray emission.
The corresponding model spectra are shown
with black solid lines in Fig.~\ref{MW_photopion} and \ref{MW_psyn}, respectively.
In the LH$\pi$ model, 
which is also defined by the higher proton injection compactness (see Table~\ref{table1}),
the TeV data are fitted by the synchrotron radiation of electron/positron
pairs that result both from charged pion decay and from the absorption of
$\gamma-$rays from neutral pion decay.  
The combination of a low magnetic field with a high proton injection 
compactness results in suppressed proton synchrotron emission
and prominent photopair and photopion components. Thus the SED does not
have the usual double hump appearance 
as synchrotron photons from the photopair 
secondaries produce a distinctive broad hump at MeV 
energies\footnote{In general we find that, 
for most relevant fitting parameters,
protons lose approximately the same amount of energy through
photopair and photopion interactions, therefore those two processes
have the same significance for the injection of secondaries.}.

In the LHs model, TeV $\gamma-$rays are produced by 
proton synchrotron radiation. The high magnetic field coupled 
with a low proton injection compactness results in a
 suppressed photopair and photopion component and two well defined peaks,
both from synchrotron radiation of electrons at low energies and
from protons at higher. 

\subsection{Neutrino emission}
In Figs.~\ref{MW_photopion} and \ref{MW_psyn}  
the neutrino spectra (light blue lines) obtained
in both models are shown along with the photon spectra. The neutrino spectral parameters, 
compared to those of other 
particle types, are summarized in Table \ref{table2}. Note that
the luminosities are given in the comoving frame.
Although the neutrino flux contains neutrinos of different 
flavours with an approximate ratio 
$ F_{\nu_e} : F_{\nu_{\mu}} : F_{\nu_{\tau}} = 2 : 1 : 0$, by the time
they reach Earth their ratio will have changed to
$ F_{\nu_e} : F_{\nu_{\mu}} : F_{\nu_{\tau}} = 1 : 1 : 1$ due
to oscillations \citep{Learned1995}. 
Since we are comparing our spectra to the IceCube
sensitivity for muon neutrinos (orange line) 
we have to scale down our results
by a factor of 3 -- see thick blue lines in Figs.~\ref{MW_photopion} and \ref{MW_psyn}.
The IC-40 limit in our figures is derived for arbitrary 
power-law neutrino spectra. Thus, for a given slope of a 
power-law neutrino spectrum the tangent to the envelope curve 
having the same slope gives the actual upper limit.

The low efficiency of photohadronic interactions in the LHs model results
in a neutrino flux that is by many orders of magnitude lower than the IC-40 upper limit and
about a factor of 10 less than the TeV $\gamma$-ray flux. The peak of the neutrino flux
occurs also at energies of 100 PeV due to the high values of the magnetic field
and the maximum  Lorentz factor of protons. 

On the other hand, the LH$\pi$ model produces a substantial neutrino flux
that is of the same order as the TeV $\gamma-$rays.
We can then see that the expected neutrino flux is
just under the sensitivity of the IC-40 detector, and should
be producing observable neutrinos for subsequent layouts with 
79 or 86 strings. The peak neutrino flux is observed at 
energies around $E_{\rm \nu,peak}= 3.3 {\rm PeV}$\footnote{Recent observations of PeV-energy neutrinos by the
IceCube collaboration \citep{IcePeV2013} are in good agreement with this
prediction, although the calculated flux is still too low to offer any
hint of a spectral shape.}.
This corresponds to an energy about 30 times
lower than the maximum proton energy in the $\nu F_\nu$ vs. $\nu$ diagram. 
At lower energies it 
follows an approximate power law with index $p_{\rm \nu}$, which is
harder than the power law of the initial protons by a factor of 
$\sim 1$,
in accordance with the approximate relation 
$p_{\rm \nu} \simeq (p_{\rm p}-0.5)/2.5$ from \citep{DMPR2012}. 
Neutrinos from neutron decay peak at an energy 
two orders of magnitude lower than those from meson decay\footnote{
The maximum neutrino energy in the neutron rest frame is
$E_{\nu,max}^{\prime n decay}=0.77MeV$, which is easily calculated
when one assumes that the produced electron is stationary. 
Halving it to obtain the mean energy and transforming to the lab frame
we get $E_{\nu,mean}^{n decay}=4.1 \cdotp 10^{-4} E_n$.}
 and their
luminosity is similarly lower. In this case their
contribution is noticeable as a small peak to the left of the main
neutrino spectrum peak in Fig. \ref{MW_photopion}. 

\begin{table}
\centering
\begin{tabular}{l  c c}
 \hline 
Parameter & model LH$\pi$ & model LHs  \\
\hline \hline

$\gpmx$ & $3.2 \times 10^6$ & $6.3 \times 10^9$ \\
$E_{\rm \nu,peak}$ & $ 1.3 \times 10^5$ & $ 1.6 \times 10^8$ \\
\hline

$p_{\rm p}$ & 1.3 & 1.5 \\
$p_{\rm \nu}$ & ~0.3 & ~0.3 \\
\hline

$L_{\rm p} (erg \ s^{-1})$ & $5.7 \times 10^{45}$ 
& $4.5 \times 10^{41}$\\
$L_{\rm e}^{inj} (erg \ s^{-1})$ & $4.6 \times 10^{40}$ 
& $ 1.2 \times 10^{41}$\\
$L_{\rm \gamma} (erg \ s^{-1})$ & $ 6.9 \times 10^{40}$ 
& $1.3 \times 10^{41}$\\
$L_{\rm \nu} (erg \ s^{-1})$ & $7.9 \times 10^{39}$ 
& $7.4 \times 10^{38}$\\
$L_{\rm n} (erg \ s^{-1})$ & $ 3.8 \times 10^{40}$ 
& $2.6 \times 10^{39}$\\
\hline

$u_{\rm p} (erg \ cm^{-3})$ & $1.6 \times 10^3$ 
& $9.7 \times 10^{-2}$\\
$u_{\rm B} (erg \ cm^{-3})$ & 1 & 100\\

$P_{\rm jet}^{\rm obs} (erg \ s^{-1})$ & $1.1 \times 10^{48}$ 
& $4.2 \times 10^{46}$\\
\hline

\end{tabular}
\caption{
Neutrino parameters compared to the respective ones of the parent proton 
distribution. The values of the proton ($u_{\rm p}$) and magnetic field ($u_{\rm B}$) 
energy densities refer to  steady state;
the luminosities of primary electrons, photons and neutrons are also included as is the total observed luminosity of the jet.
}
\label{table2}
\end{table}

\subsection{UHECR emission}

\begin{figure}
\centering
\resizebox{\hsize}{!}{\includegraphics{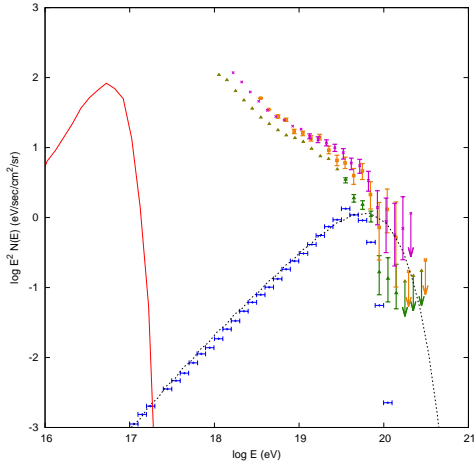}}
\caption{\small{
UHE proton spectra resulting from neutron decay and obtained within the LH$\pi$ (red line) and
LHs (black dashed line) models. For the latter, the UHE cosmic ray spectra obtained after taking into account
propagation effects using the numerical code CRPropa 2.0 \cite{Kampert2013} are also shown (blue crosses). 
The UHE cosmic ray energy spectrum as observed by
Auger \citep{PA2011}, HiRes-I \citep{HiRes2009} and Telescope Array 
   \citep{TA2013} is overplotted 
   with green open triangles, magenta open squares and orange x's
    respectively. 
}}
\label{pn}
\end{figure}

Neutrons resulting from photopion interactions are an effective
means of facilitating proton escape from the system, as they are unaffected
by its magnetic field and their decay time is high enough to allow
them to escape freely before reverting to protons 
\citep{KM1989, Begelman1990, Giovanoni1990, AtoyanDermer2003}. 
A further
advantage is that they are unaffected by adiabatic energy losses
that the protons may sustain in the system as it expands
\citep{Rachen1998}. Those effects make them excellent originators
of UHECR.
In Fig.~\ref{pn} we show the ultra-high energy neutron spectrum 
having first decayed into protons, as 
obtained by the LH$\pi$ (red solid line) and the LHs (black dashed line) models.

The plotted spectrum of the LH$\pi$ model is just an upper limit 
of what it would appear at Earth, in the absence of cosmic-ray diffusion. We have not 
taken that into account because 
treating it at energies $<10^{17}$ eV would lie
outside the scope of the present paper that focuses on UHECR. At any rate, our values, even as an upper limit,
 are well below the observed CR flux at such energies. 

In the LHs model, however, the higher value of the maximum proton Lorentz factor
used in the SED fitting makes the discussion 
about UHECR emission more relevant -- see black dashed line in Fig.~\ref{pn}. 
In this case, the propagation of UHE protons in a uniform intergalactic ${\rm pG}$
magnetic field (transverse to the direction of propagation) and their energy losses from interactions with the 
cosmic microwave and infrared and optical (IRO) backgrounds 
were modelled using CRPropa 2.0 \citep{Kampert2013}. In particular, the mean free paths of UHE protons
for pair- and pion-production processes on the IRO backgrounds are modelled using the best-fit model by \cite{kneiske04}.
Since at this energy range those losses are the dominant factor in 
reshaping the proton spectrum and diffusion effects are minimal,
we have restricted ourselves to one-dimensional (1D) simulations.
The resulting spectra in Fig.~\ref{pn}
(blue crosses) are compared to 
observations from Auger \citep{PA2011} (open triangles), 
HiRes-I \citep{HiRes2009} (open squares) 
and Telescope Array \citep{TA2013} (x-symbols). 
At energies $30-60$ EeV the expected cosmic rays are just
under the present limits; we remind that our results are 
obtained by fitting Mrk~421 in a low state. During a flaring
state the produced UHECR flux will be even higher\footnote{It is worth mentioning that during the time
of writing Mrk~421 underwent a large flaring event seen in optical
(private communication with Dr. K. Gazeas),
X-rays \cite{kapanadze13}, GeV
 \cite{paneque13} and TeV $\gamma$-rays \cite{cortina13}.}. 
This could explain some of
the discrepancy between the Auger and HiRes--I/Telescope Array data 
 at that energy range, since
Mrk~421 is in the northern sky and, thus, invisible to Auger. 
Interestingly enough, the same
applies for every other high-frequency-peaked BL Lac within a distance 
of $z=0.05$ \citep{Senturk2013}.
It is more difficult to assess the contribution of escaping protons
to the UHECR spectrum. The proton flux is two orders of magnitude higher 
than the neutron one but, unlike neutrons, 
protons are subjected to adiabatic losses after escaping from
the active region. In particular, if the emitting region 
lies within an expanding jet, the escaping protons will be susceptible to
adiabatic energy losses and may end up carrying 
a negligible fraction of the UHECR flux, while
the main contribution comes from protons produced by neutron decay. 
A study of the escaping proton 
contribution lies outside the scope of the present paper
since it requires assumptions on the specific geometry of the jet. 
 
\begin{figure}
 \centering
\resizebox{\hsize}{!}{\includegraphics{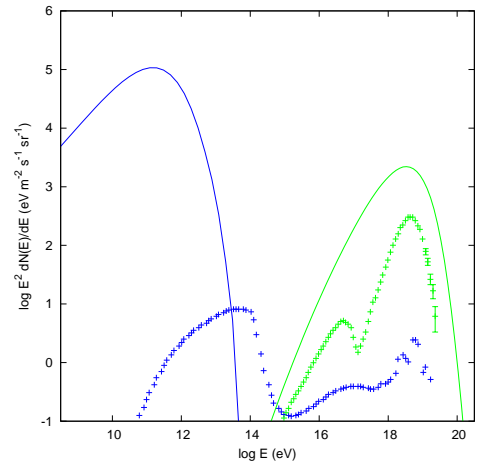}}
\caption{
Photon (blue solid line) and neutrino of all flavours (green solid line)  spectra according to the LHs model. 
 The secondary $\gamma-$ray and neutrino spectra resulting from the UHE proton propagation are also shown with blue and green
 crosses, respectively.
The propagation of the UHE protons into a uniform pG intergalactic magnetic field was modelled using CRPropa 2.0.
}
\label{comb}
\end{figure}

An interesting question that might be addressed is the level of the line-of-sight
neutrino/$\gamma$-ray flux resulting from the UHE proton propagation. 
Figure \ref{comb} shows the secondary $\gamma-$ray (blue crosses) and neutrino spectra (green crosses) produced by interactions of UHE protons
with the CMB and IRO backgrounds, derived for the LHs model\footnote{The
low energies  ($< 0.1$ EeV) of the cosmic rays produced within the LH$\pi$ model made an analogous calculation
using the numerical code CRPropa 2.0 not possible.} using CRPRopa 2.0;
our model
spectra coming directly from the source (blue and green solid lines)  are overplotted for comparison reasons.
We note that in 1D propagation simulations the strength of the uniform intergalactic magnetic
field affects only the synchrotron emission of the secondaries produced during the UHECR propagation while it does not
cause any deflections; in this sense, it can be considered as an upper limit. 
 Even in the extreme case of an intergalactic magnetic field of 1 $\mu$G strength, the secondary $\gamma-$ray and neutrino
emission was found to be at least one order of magnitude below the LHs  source emission.
The assumption of a structured intergalactic magnetic field would result in the deflection of UHECR by an angle of a few degrees
(see e.g. \citep{achterberg99}) with respect to our line-of-sight and, therefore, lead to an even lower flux emitted by the charged secondary particles produced during
the propagation.

\section{Summary/Discussion}
\label{discussion}

In the present paper we have calculated the detailed neutrino 
and neutron spectral flux
emerging from the BL Lac object Mrk~421 under the assumption of
a one-zone leptohadronic model. 
Even if the assumed geometry of the source is
simple enough, we have applied a sophisticated method to calculate in a
self-consistent and time-dependent way the photopair and photopion interactions.
For this we used the numerical code
presented in \citep{DMPR2012} and made fits to the MW spectrum 
of this source
as obtained in 2001 using contemporaneous optical, X-ray and TeV
$\gamma-$ray data \citep{Fossati2008}. Using then the so-obtained fitting parameters
allowed us to calculate the expected neutron and neutrino fluxes.
Consequently we have followed the propagation of protons, which result
from the escaping neutron decay, and thus calculated the expected 
UHECR contribution
of Mrk~421 at Earth.

We found that two sets of very different parameters produce good
fits to the SED of the source. 
In the first case, $\gamma-$rays are produced from synchrotron radiation 
of secondaries resulting from
photopion interactions, therefore they are pion-induced (the LH$\pi$ model).
In the other case, the $\gamma-$rays originate from
proton synchrotron radiation (the LHs model).

Proton acceleration at  ultra-high energies, i.e. $E_{\rm p,max}= 1.4 \times 10^{20}$ eV,
is a typical feature of the LHs model (see also \cite{Mucke2001}),
 which agrees well with the requirement for UHECR accelerators.
Although it is magnetically dominated 
it is a much more
economic model than LH$\pi$ in terms of total jet luminosity -- see Table~2.
Propagation of the protons produced from the escaping neutron decay
results in a UHECR flux at Earth that is very close to the measurements
of current experiments
at energies around 30 EeV. However, due to the fact that
our UHECR spectrum is peaked at high energies, its overall shape is very different from the observed one 
at energies below 30 EeV -- see Fig.~\ref{pn}. 
Even if one assumes that all other Northern Hemisphere nearby BL Lac objects 
produce the same spectral shape  of UHECRs as Mrk~421 and normalize their cosmic-ray output
to their photon luminosity, their contribution to the total UHECR flux will not
be significant because of the combination of their low luminosities and of the cosmic-ray propagation  through larger distances;
 we remind that among these sources, Mrk~421 is not only the closest blazar but also the most
luminous one.
Furthermore, the LHs model produces a low neutrino flux, since photohadronic
processes are suppressed to a level that is many orders of magnitude
below the IceCube sensitivity threshold -- see Fig.~\ref{MW_psyn}. 
While it is possible to fit the SED with steeper proton injection spectra,
we found that these cases cannot alter significantly our conclusions regarding UHECR and neutrino fluxes,
as long as $p_{\rm p} < 2.5$.

The LH$\pi$ model, on the other hand, requires a large, but not unacceptable,
jet power which is heavily particle dominated -- see Table 2. Good fits to the
SED of the source are obtained assuming that the protons are accelerated up to
$E_{\rm p} \simeq 30$ PeV, therefore they cannot contribute to the UHECR flux. 
However, the produced neutrino
flux is tantalizingly close to the IC-40 sensitivity limit for Mrk~421 \citep{Tchernin2013} and 
peaks within the energy range where the first PeV neutrino detection from the IceCube collaboration was
reported \citep{IcePeV2013}.  A shortcoming of the LH$\pi$ model is its low efficiency.
If we define the total efficiency $\xi$ as the ratio of 
the sum of escaping luminosities in photons, neutrons and neutrinos 
over the sum of the injected luminosities, then
we find $\xi \approx 2\times 10^{-5}$. That could be partially alleviated 
if we assume that the protons are injected at a lower luminosity but gradually pile
up inside the source due to a larger escape time compared 
to the crossing time of the source. 

The present work is focused on the high-peaked blazar 
Mrk~421. 
Thus, the distribution of ultra-high energy protons 
that we use in order to derive the neutrino and UHECR spectra
is specifically determined by fitting its contemporaneous SED. 
Generally, we find that 
homogeneous leptohadronic one-zone models can give very good fits to the
MW observations of Mrk~421. 
A question that
arises then is whether such results can be 
safely generalized and, if so,
to what extent. For example, in our calculations we used only 
intrinsically produced photons as targets 
for the photohadronic interactions; this can be safely assumed in the case
of high-peaked BL Lacs but by no means can be generalized. 
Thus, if we were to model another source, e.g. 
a flat spectrum radio quasar which is generally more luminous than a typical BL Lac,
and has a high abundance of external
photons (e.g. from the Broad Line Region), we 
would require very different parameters from the ones derived here.

An approach for calculating diffuse neutrino and UHECR fluxes would 
have then been
to calculate simultaneous MW fits to the blazar sequence 
\cite{ghisellinietal98}
by means of a leptohadronic model whenever possible.
Using these as templates one could, in principle, calculate the neutrino
and neutron
flux for each class of objects and then integrate over redshift. We plan
to perform
this type of calculation in a forthcoming paper.

\section*{Acknowledgements}
This research has been co-financed by the European 
Union (European Social Fund – ESF) and Greek national 
funds through the Operational Program "Education and 
Lifelong Learning" of the National 
Strategic Reference Framework (NSRF) - Research Funding 
Program: Heracleitus II. Investing in knowledge society through the European Social Fund. 
We would like to thank the
anonymous referee for useful comments on the manuscript.

\end{document}